\documentclass[]{ceurart}
\usepackage{listings}
\usepackage{amsmath}
\usepackage{booktabs}
\usepackage{tikz}
\usetikzlibrary{shapes.geometric, shapes.multipart, arrows.meta, positioning, calc, backgrounds, fit}
\usepackage{standalone}
\usepackage{cleveref}

\definecolor{kgBlueBase}{HTML}{2A78D6}
\definecolor{kgOrangeBase}{HTML}{EB6834}

\colorlet{kgArtLine}{kgBlueBase!90!black}   %
\colorlet{kgArtFill}{kgBlueBase!7}          %
\colorlet{kgSwLine}{kgOrangeBase!80!black}  %
\colorlet{kgSwFill}{kgOrangeBase!7}         %
\colorlet{kgModel}{black!58}                %

\providecommand{\gelab}[1]{{\color{kgArtLine}\sffamily #1}}
\providecommand{\gelabsw}[1]{{\color{kgSwLine}\sffamily #1}}
\providecommand{\gelabmod}[1]{{\color{kgModel}\sffamily #1}}
\providecommand{\kt}[1]{{\normalfont\scriptsize\itshape\color{black!57}#1}}
\providecommand{\mo}[1]{{\color{kgModel}#1}}

\usetikzlibrary{positioning, calc}

\makeatletter
\@ifundefined{vgap}{\newlength{\vgap}}{}
\@ifundefined{hgap}{\newlength{\hgap}}{}
\makeatother
\tikzset{
  kgLab/.style={
    text centered, fill=white, opacity=0.98, text opacity=1,
    inner sep=0.2ex, font=\sffamily\footnotesize
  },
  kgIri/.style={
    draw=black!50!white, rectangle, rounded corners, thick,
    text centered, top color=white, bottom color=black!15,
    font=\sffamily\footnotesize
  },
  kgArtNode/.style={kgIri, draw=kgArtLine, top color=white, bottom color=kgArtFill},
  kgSwNode/.style={kgIri, draw=kgSwLine, top color=white, bottom color=kgSwFill},
  kgArr/.style={thick, font=\sffamily\footnotesize},
  kgArrOut/.style={kgArr, -latex, pos=0.45},
  kgArrIn/.style={kgArr, latex-, pos=0.55},
  kgArrInOut/.style={kgArr, latex-latex}
}

\usetikzlibrary{shapes.multipart}
\tikzset{
  kgClassNode/.style={
    rectangle split, rounded corners, thick, draw=black!50!white,
    align=left, font=\sffamily\footnotesize, inner sep=7pt
  },
  kgArtClassNode/.style={
    kgClassNode, rectangle split parts=2, draw=kgArtLine,
    rectangle split part fill={kgArtFill!60, white}
  },
  kgSwClassNode/.style={
    kgClassNode, rectangle split parts=3, draw=kgSwLine,
    rectangle split part fill={kgSwFill!60, white, white}
  }
}

\begin{document}

\copyrightyear{2026}
\copyrightclause{Copyright for this paper by its authors. Use permitted under Creative Commons License Attribution 4.0 International (CC BY 4.0).}

\title{From Code Archival to Knowledge Graph: Bridging Software Heritage, COAR Notify and Wikidata}

\author[1]{Camillo Carlo {Pellizzari di San Girolamo}}[orcid=0000-0003-2699-1693, url=https://www.sns.it/en/persona/camillo-carlo-pellizzari-di-san-girolamo, email=camillo.pellizzaridisangirolamo@sns.it]
\address[1]{Scuola Normale Superiore, p.zza dei Cavalieri 7, 56126 Pisa PI, Italy}

\author[2]{Francesco Tosoni}[orcid=0000-0001-8457-3866, url=https://www.santannapisa.it/en/francesco-tosoni, email=Francesco.Tosoni@santannapisa.it]
\cormark[1]
\address[2]{Sant'Anna School of Advanced Studies, L'EMbeDS, p.zza Martiri della Libert\`a 33, 56127 Pisa PI, Italy}

\cortext[1]{Corresponding author.}

\conference{Wikidata'26: Wikidata Workshop at ISWC 2026}

\begin{abstract}
  Software is a first-class scientific object, yet validated links between source code and the scholarly record remain largely absent from the Linked Open Data (LOD) cloud, isolating archived artefacts from semantic discovery. This paper presents an end-to-end reconciliation pipeline that harvests, validates, and models publication-to-repository pairs from sources where the link between a paper and its source code is explicit and editorially verified: the software-centric journals JOSS, SoftwareX, and IPOL, together with the reproducibility reports of the SIGMOD Availability and Reproducibility Initiative (ARI). This yields a curated corpus of 4,397 $\langle$DOI, repository-URL$\rangle$ pairs. We design two distinct application profiles grounded in Wikidata classes (one for scholarly articles, one for software instances) aligned with the schema.org and CodeMeta vocabularies. This architectural separation enables rule-based reconciliation at two granularities: lightweight, inline publication references or standalone, first-class Wikidata software nodes equipped with SWHIDs, Software Heritage's content-addressed identifiers. A read-only lookup against Wikidata shows that only 82 of the harvested repositories were already modelled there; human-reviewed batches have since created 4{,}182 new software items cross-linked to their articles. We further show that payloads of the emerging COAR Notify protocol, an external effort we do not develop, map natively onto our input format, so the same backend could later serve a live enrichment stream. Our core contribution is a pair of application profiles that turn Wikidata into a connector between the scholarly record and archived source code; we openly release all code, application profiles, and harvested datasets.
\end{abstract}


\maketitle

\section{Introduction}
\label{sec:intro}

Reproducibility and discoverability are foundational to science, yet the transition to software pipelines has eroded both: investigators fail to validate roughly 70\% of published results, and about half cannot reproduce their own findings~\cite{greengard2026repro}. A primary cause is lost or unobtainable source code. As Karl Popper observed \cite[\S22]{popper1959logic}, ``Non-reproducible single occurrences are of no significance to science.'' Durable, unambiguous software identification is therefore a prerequisite for scientific trust, requiring code to survive and remain linked to the scholarly record.

Although software encodes modern scientific methodology, it is rarely cited as a formal, resolvable object~\cite{strasser2022rules,dicosmo2020swh}. Two complementary infrastructures address this gap. Software Heritage (SWH) \cite{swh-ecosystems-book-chapter,swh_cacm,dicosmo2020swh} archives public source code and assigns persistent, content-addressed identifiers (SWHIDs) via a Merkle directed acyclic graph (DAG)~\cite{swh_dag}. Meanwhile, Wikidata \cite{wikidata-survey,wikidata-cacm,wikidata-making} (the central knowledge base of the Linked Open Data cloud) models software items alongside metadata such as DOIs and SWHIDs.

Each infrastructure solves half the problem: SWH guarantees preservation, while Wikidata provides a queryable semantic layer connecting software to papers, authors, and licences. Systematic bridging is missing. Connecting these graphs enriches discovery, enabling queries by functional properties (e.g., language, licence, topic, or author) rather than solely by URL or hash, which also makes such data reusable by developers outside the knowledge-graph community~\cite[\S5.7]{kg_role_dagstuhl}.

Building this bridge aligns with research-software policy: funders ask that software be treated as a first-class scholarly output and preserved in archives such as SWH~\cite{strasser2022rules}, and research-software practitioners call for interconnected portals that integrate software into the scholarly record~\cite{dicosmo2025roadmap}, alongside the scholarly graph that WikiCite has grown in Wikidata~\cite{wikicite_report_2016}. In France, the IPOL journal already deposits its code in SWH~\cite{ipol-swh} and national prizes reward research software~\cite{dicosmo2025roadmap,code-beyond-fair}.

The scale of this semantic gap is stark. As of July 2026, Wikidata's code-repository property (\texttt{P1324}) appears on 22,534 items, but only 133 include a DOI (\texttt{P356}), and only 65 carry both a SWHID (\texttt{P6138}) and a repository link. Although the ontological scaffolding exists, it remains sparsely populated. This paper designs and prototypes the reconciliation machinery required to fill it.

Beyond provenance, content-addressed links from papers to archived code let users navigate to runnable code, study software empirically at scale, and ground code search engines (e.g., MediaWiki Code2Code Search~\cite{tosoni2026mediawiki}) and code models (such as StarCoder2~\cite{the-stack-v2}) in the scholarly record, complementing bibliographic graphs like DBLP~\cite{dblp_lessons}.

\paragraph{Our contribution}

Our primary outcomes are fourfold: (i) we consolidate editorially verified software-journal and reproducibility-report sources into an open corpus of 4{,}397 curated $\langle$DOI, repository$\rangle$ pairs whose paper-to-code link is correct by construction; (ii) as our central contribution, we model software-citation links using separate application profiles for scholarly articles and software instances aligned with schema.org \cite{schema_org} and CodeMeta \cite{codemeta}, turning Wikidata into a connector that reconciles SWH with the scholarly record at two granularities; (iii) we present a rule-based reconciliation pipeline that generates auditable, provenance-tagged candidate statements for human review; (iv) we show that a payload of the emerging COAR Notify protocol \cite{coarnotify} maps directly onto our pipeline's input format, so the offline prototype and a future live notification stream could share a single processing backend.

\section{Background}
\label{sec:background}

\subsection{Software Heritage}
\label{sec:swh}

\begin{figure}
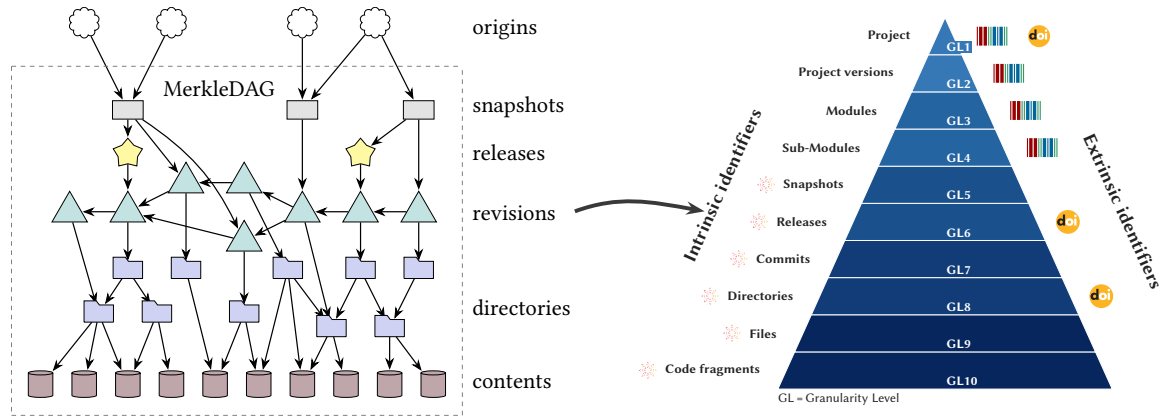

  \centering
  \resizebox{\linewidth}{!}{\begin{tikzpicture}

  \node[inner sep=0pt, anchor=north] (dag) at (0,0) {%
    \resizebox{!}{11cm}{\input{figs/swh-datamodel.tex}}%
  };

  \node[inner sep=0pt, anchor=north, right=1.2cm of dag] (pyr) {%
    \input{figs/pyramid.tex}%
  };

  \draw[-{Stealth[length=14pt, width=10pt]}, line width=3pt, gray!45!black]
    ([xshift=3pt]dag.east) to[out=20, in=170] ([xshift=2.4cm]pyr.west);

\end{tikzpicture}}
  \caption{Granularity levels of the Software Heritage Merkle DAG \cite{swh-ecosystems-book-chapter} and corresponding identifier schemes \cite{gruenpeter2026digital}. \emph{Intrinsic} SWHIDs address content-defined layers (from snapshots and releases down to code fragments) and are derived from the objects themselves. \emph{Extrinsic} identifiers minted by external authorities attach wherever a human-meaningful unit is deposited and cited: typically a project, version or release, but also a directory when an archive mints a DOI for a deposited source tree. Property \texttt{P6138} allows a project-level Wikidata item to reference its underlying intrinsic SWHID anchor.}
  \label{fig:granularity}
\end{figure}

Software Heritage (SWH) \cite{swh-ecosystems-book-chapter} was launched in 2016 to collect, preserve, and share all publicly available software source code~\cite{dicosmo_why_how,swh_cacm}. As of December 2025, it archives more than 27 billion unique source files from over 421 million projects, totalling roughly 2 petabytes~\cite{swh_report_2025}.

SWH occupies a distinct place among code platforms. Forges such as GitHub or GitLab are commercial services that host active development, where projects can be renamed, moved, or deleted at will, while deposit repositories such as Zenodo mint a DOI for a snapshot that authors upload themselves. SWH is instead a non-profit archive, initiated by Inria in partnership with UNESCO and funded by public and private sponsors~\cite{dicosmo_why_how,swh_cacm}: it hosts no development, but proactively crawls forges and package managers and preserves their full history. Replicated to independent mirrors, any archived artefact can be retrieved, redistributed, and cited through its intrinsic identifier wherever it was first published, so the code behind a result stays verifiable without depending on a single provider.

The archive is structured as a five-level Merkle DAG \cite{swh_cacm,swh_dag,dicosmo2020swh} representing contents, directories, revisions, releases, and snapshots, rooted at origins~\cite{swh_dag}. Because the graph is content-addressed, identical files collapse into a single node shared across all projects, enabling global deduplication and a stable identity model. This design ensures resilience against hosting platform closures: when Gitorious and Google Code shut down, SWH retrieved and preserved full copies of their repositories~\cite{dicosmo_why_how,dicosmo2020swh}.\footnote{See Software Heritage, \emph{Software is fragile}: \url{https://www.softwareheritage.org/mission/software-is-fragile/}.}

Every archived object receives a SWHID (SoftWare Hash IDentifier), an intrinsic identifier derived directly from its content, rendering it independent of hosting platforms and immune to link rot~\cite{dicosmo2020swh,di-cosmo-archiving}. SWHIDs are codified under the ISO/IEC 18670:2025 standard \cite{ISO-IEC-18670-2025} and supported via the Wikidata property \texttt{P6138}. \Cref{fig:granularity} contrasts the intrinsic SWHID with extrinsic identifiers, such as Wikidata Q-items and DOIs (including identifier systems like RRID \cite{rrid2016initiative}), which attach at coarser, human-meaningful levels, such as projects or releases. The archive also supports research on programming-language evolution~\cite{SWH-filepath-assignment}, code language models trained on The Stack v2~\cite{the-stack-v2}, reproducible deployment~\cite{zacchiroli-archiving-repro}, and cybersecurity~\cite{SWH-cybersecurity}, and the UNESCO Recommendation on Open Science~\cite{UNESCO2021OpenScience} lists open source software and source code among the pillars of open scientific knowledge.

\subsection{Wikidata}
\label{sec:wikidata}

Wikidata is a collaborative, multilingual knowledge base serving as the central hub of the Linked Open Data cloud~\cite{wikidata-survey,wikidata-cacm,wikidata-making}. It is structured into subject-property-value triples, which can be qualified and equipped with references to track provenance. This suits software metadata, which evolves across versions, although curating such a graph remains challenging~\cite{dblp_lessons}\cite[\S3.8]{kg_role_dagstuhl}.

Software items in Wikidata are modelled as instances of \emph{software} (\texttt{Q7397}) and its subclasses, using established properties such as source code repository (\texttt{P1324}), DOI (\texttt{P356}), and SWHID (\texttt{P6138}).

This article-versus-software distinction is reflected in Wikidata's physical infrastructure. In 2025, the Wikimedia Foundation split the Wikidata Query Service (WDQS) into two SPARQL endpoints to overcome backend scaling limits~\cite{wdqs-split}. One endpoint serves the scholarly graph, which is primarily WikiCite data~\cite{wikicite_report_2016} and accounts for over half of Wikidata's triples~\cite{wdqs-split}, while the main endpoint hosts software items (\texttt{Q7397}). As of July 2026, queries on WDQS that join a paper to its corresponding software must still span both graphs via internal federation, an architectural constraint we address in \Cref{sec:granularity}.

\subsection{COAR Notify and the Reconciliation Trigger}
\label{sec:coar}

\begin{figure}
  \centering
  \resizebox{\linewidth}{!}{\input{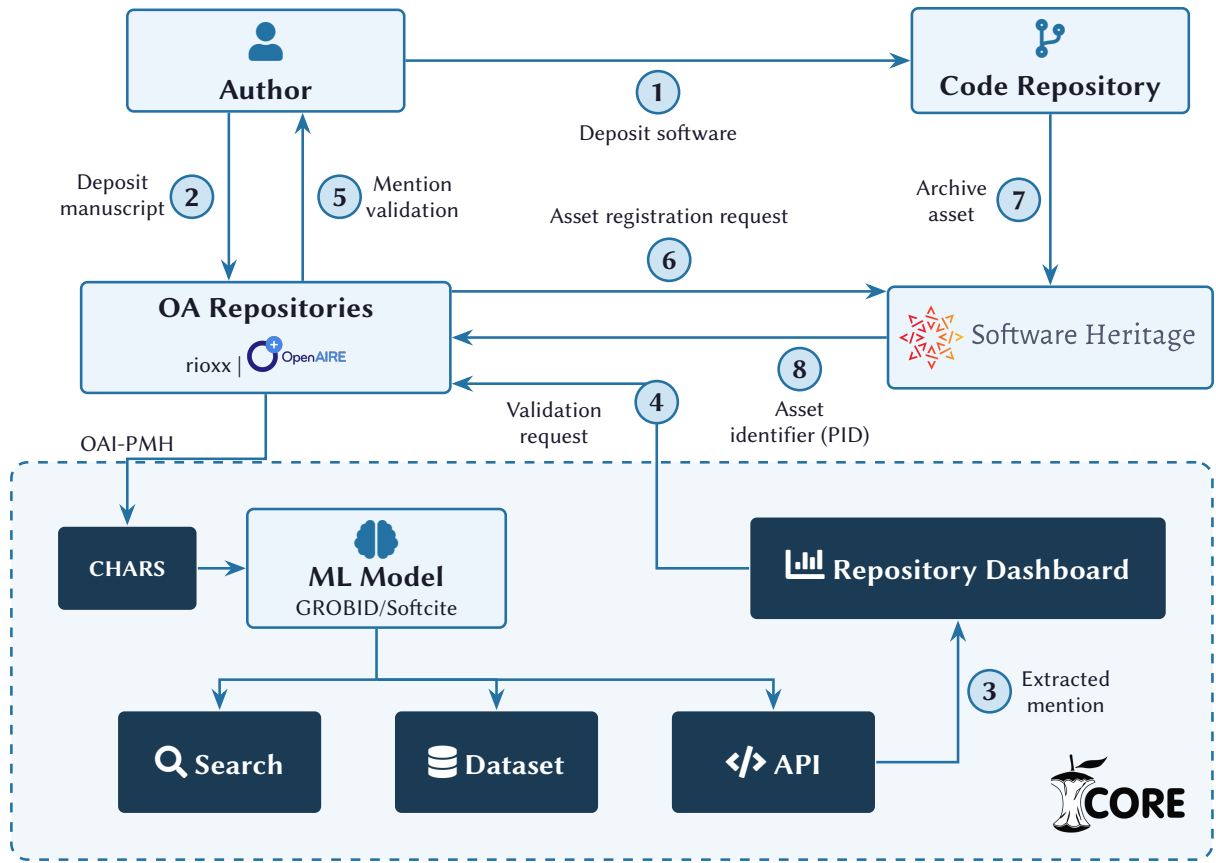}}
  \caption{COAR Notify ecosystem for software-citation discovery and archival, after the SoFAIR workflow of Cancellieri et al.~\cite{coar} (numbered arrows indicate workflow steps). Authors deposit code to a hosting platform and the corresponding manuscript to an OA repository (1--2); CORE harvests the deposit over OAI-PMH and extracts software mentions via its ML pipeline (3); the Repository Dashboard then routes a validation request to the repository (4), which forwards it to the author (5); once validated, the repository registers the asset with Software Heritage, which archives it and returns its persistent identifier (6--8).}
  \label{fig:coar-notify-model}
\end{figure}

The COAR Notify protocol defines machine-actionable notifications, expressed via Activity Streams 2.0, for scholarly communication workflows~\cite{coarnotify,coar}. The COAR community develops it; Cancellieri et al.~\cite{coar} have blueprinted its application to software-citation discovery at scale as part of the SoFAIR project, but have not yet deployed it operationally. We neither develop this protocol nor operate its notification infrastructure: we treat it strictly as an external, prospective input channel and show only that its payloads align with what our pipeline already consumes.

Open Access (OA) repositories expose their deposits via OAI-PMH conforming to the RIOXX profile~\cite{rioxx}, an application profile that extends Dublin Core with structured fields for funder identifiers, licensing terms, and author ORCIDs. This common representation lets CORE harvest both the deposited PDFs and their provenance metadata: CHARS downloads the PDFs;\footnote{Component names follow the SoFAIR workflow documentation, \url{https://sofairoa.github.io/documentation}.} the GROBID suite \cite{grobid2009,grobid2020evaluation} parses each into structured XML (TEI) representations; Softcite isolates software mentions from that structure; and the Repository Dashboard sends a validation request back to the originating OA repository as a COAR Notify message, which the repository forwards to the author for approval (arrows 3 to 5 in \Cref{fig:coar-notify-model}).

In our architecture, each COAR Notify \texttt{Announce} payload carries a paper's DOI alongside the repository URL of related software, mapping directly to the $\langle$DOI, repo$\rangle$ pairs consumed by our reconciliation pipeline (\Cref{sec:methodology}). These pairs are then reconciled against Wikidata and serialised as QuickStatements \cite{quickstatements} to mint or enrich software items with repository links (\texttt{P1324}) and SWHIDs (\texttt{P6138}), at node or reference level depending on whether the paper presents or merely uses the software (\Cref{sec:granularity}), making COAR Notify a natural on-ramp for a future live enrichment stream.

\section{Datasets: Original Sources and Derived Pairs}
\label{sec:datasets}

The systematic survey of entity linking with Wikidata by Scharpf et al.~\cite{wikidata-survey} evaluates 65 publications, 17 benchmark datasets, and 34 approaches. Crucially for our purposes, the only ones it describes as covering software mentions in scientific literature are Softcite~\cite{softcite2021} and SoMeSci~\cite{somesci2021}. Both are mention-level corpora: they annotate where a paper refers to third-party software so a single article can point to many tools, and they recover repository URLs from free text, which makes them broad in coverage but noisy. This scarcity and noise motivate building the corpus instead from sources where the paper-to-code link is explicit and editorially verified. Preprint full text is a complementary avenue: arXiv distributes the \LaTeX{} sources of its papers, and corpora such as unarXive~\cite{saier2020unarxive} convert them into over a million plain-text documents. Mining that text for availability statements (patterns such as ``code is available at'' followed by a URL) would raise recall at the cost of editorial verification; we leave that cross-check to future work.

We distinguish \emph{original} datasets, the bibliographic sources we consume, from \emph{generated} datasets, the $\langle$DOI, repo$\rangle$ pair tables our harvester derives from them (\Cref{tab:generated}). In our sources, the editorial workflow ties each publication to a repository, so the link is correct by construction rather than inferred from prose.

\subsection{Software-Journal Ground Truth}
\label{sec:groundtruth}

We gathered repository links for the three software journals using a tiered, API-first strategy. We queried the JOSS JSON catalogue directly (3{,}596 records covering 3{,}546 distinct works, with only one lacking a repository) and covered SoftwareX via Crossref API queries by ISSN (2{,}326 works). Because IPOL lacks a machine-readable endpoint, we fetched its landing pages once and cached them locally, then extracted the SWHID from the exposed BibTeX records (295 papers). After URL normalisation, SWHID resolution, and deduplication against the rest of the corpus, these venues contribute 3{,}545, 486, and 275 pairs, respectively, to the unified table (\Cref{tab:generated}).

\paragraph{SoftwareX (Elsevier, ISSN 2352-7110)} 

This journal requires a public code repository for every article, and Elsevier mirrors each accepted repository into a dedicated \texttt{ElsevierSoftwareX} GitHub organisation, nominally one mirror per paper. However, the article's Crossref metadata does not expose the repository link. We therefore harvested 2{,}326 SoftwareX DOIs from Crossref by ISSN, each carrying an Elsevier Publisher Item Identifier (PII), and independently enumerated the organisation's mirror repositories. We then recovered the paper-to-repository link through the PII embedded in each mirror's GitHub description, which serves as the join key back to the Crossref record, with a normalised-title match as a fallback.\footnote{This join trusts the PII that Elsevier records in each mirror's description. We found one mirror, \texttt{ElsevierSoftwareX/SOFTX\_2020\_261} (upstream \texttt{PhotonSTR-18}), whose description cites the PII of an unrelated article (\emph{cashocs}, DOI \texttt{10.1016/j.softx.2020.100646}) instead of its own (DOI \texttt{10.1016/j.softx.2020.100640}), yielding a spurious pairing. It surfaced as a duplicate DOI and was excluded from the import. We found no further collisions of this kind in the venue.} For the \texttt{P1324} value, we record the mirror's upstream parent (the authors' original repository) rather than the Elsevier mirror itself; of the 665 mirrors matched to an article, only the 486 exposing such a parent enter the corpus. In contrast, we retain the mirror URL as the provenance reference for the mapping.

\paragraph{The Journal of Open Source Software (JOSS, ISSN 2475-9066)}

This venue represents an even stronger paradigm~\cite{joss2018}: a JOSS paper briefly describes a software repository, where the repository URL is a mandatory submission field and peer review is conducted directly on the codebase. The mapping from a JOSS DOI to its repository is therefore one-to-one and trivially machine-extractable, making JOSS an ideal gold standard.\footnote{A few submissions nonetheless list the paper's own support repository (its last path segment being \texttt{paper} or \texttt{joss\_paper}) rather than the software; four such rows are excluded at import, mirroring the SoftwareX case above. Where the recorded link resolves to a branch view or a paper subfolder, it is trimmed to the repository root (see \Cref{sec:methodology}).}

\paragraph{The Image Processing On Line journal (IPOL, ISSN 2105-1232)}

The venue, hosted at \url{https://www.ipol.im/}, offers the most robust case. IPOL publishes image-processing algorithms alongside their reference implementations and systematically deposits their source code into Software Heritage~\cite{ipol-swh}. Because an archived article embeds its assigned SWHID directly within its exported BibTeX record, it yields a $\langle$DOI, SWHID$\rangle$ pair where the archival anchor (\texttt{P6138}) is established by design. As the only source in our study that provides native SWHIDs, IPOL bypasses both full-text URL extraction and downstream repository-to-SWHID resolution, thereby yielding an ideal seed dataset for node-level reconciliation.\footnote{Because IPOL exposes no repository, the SWHID is the item's sole archival anchor. Of the 275 harvested pairs, 98 carry neither a repository nor a resolvable SWHID and are excluded at import, since the resulting node would have no anchor at all; the remaining 177 are imported as software items bearing a title-derived label, an \texttt{instance of} statement, the SWHID (\texttt{P6138}), and the cross-link to their article, with no \texttt{P1324}. Three SWHIDs harvested from the landing-page BibTeX were malformed, concatenating several archived snapshots from a single origin; we repaired each to the single authoritative qualified SWHID resolved via the Software Heritage API. One of these, \texttt{10.5201/ipol.2011.cm\_fds} (\emph{Finite Difference Schemes for MCM and AMSS}) in fact bundles two distinct implementations archived under separate SWH origins; following the multi-repository pattern of \Cref{sec:linking}, we model it as a single article with two software items. Conversely, the two DOIs of \emph{Image Interpolation with Geometric Contour Stencils} (\texttt{g\_igcs}, 2011 and 2012) share one SWHID and hence one software item. IPOL thus contributes 177 software items across 178 import rows.}

\paragraph{The SIGMOD Availability and Reproducibility Initiative (ARI)}

To test generality beyond venues specialised in research software, we include the SIGMOD ARI, which has published individual reproducibility reports for badged papers since 2020; because these reports are purpose-written, they reliably name the code repository. We scraped the ARI landing page to enumerate all badged papers for 2020--2025 (295 total); of these, only 102 carry an individual report PDF (the remaining badged papers list no standalone report). The reports for 2020--2023 are served directly by the ARI site, while 24 of the 72 badged 2024 papers have reports on the ACM Digital Library; the 2025 cohort (107 papers) links collectively to the proceedings rather than to individual reports and was therefore excluded. We extracted repository URLs by combining regex matching over the PDF text with annotation-link extraction from the PDF metadata, then ranked them by occurrence count and by the presence of artefact-signal terms (\textit{reproducib}, \textit{artefact}) in the repository path. To audit this heuristic, a large language model (Claude), blind to the regex output, named the primary artefact repository from each report's text, and we manually inspected every disagreement with the automated pick. The review found the heuristic to be correct in 90 of 102 reports (88.2\%). It improved the remaining twelve: nine corrected (chiefly URLs truncated by PDF line-wrapping or by an over-eager suffix-stripping bug, plus one case where the authors' origin repository replaced an ARI reviewer's clone) and three recovered (one repository wrongly discarded by an over-broad filter and two artefacts hosted on self-hosted GitLab instances outside the major-host allowlist). The validated pipeline yields 91 pairs from the 102 available reports (89.2\%); the eleven misses reference no permanent repository, distributing their artefacts instead through cloud drives, anonymised links, or non-public archives.\footnote{At import, an audit of the SIGMOD artefact repositories removed thirteen of the ninety-one harvested pairs. A shared reproducibility mono-repo (\texttt{damslab/reproducibility}) cited by five distinct Apache SystemDS tools was filtered out, since one repository cannot stand as the source repository of five separate software items; one further row carried no repository, SWHID, or derivable label; and seven rows were reproducibility-artefact repositories whose slug is a generic \texttt{sigmod\emph{NN}}/\texttt{reproducibility} name and whose paper names no tool from which to derive a clean label, so importing them would have minted mislabelled, low-quality items. Where the artefact repository is a reproducibility wrapper but the tool name is recoverable, the label is taken from the paper's short title (a single clean token such as \emph{Grafite} or \emph{PimPam}) or set explicitly (\emph{SQLSolver}, \emph{DivExplorer}, \emph{AU-DB}, \emph{Clonos}, \emph{Tuplex}). These exclusions take SIGMOD from 91 harvested pairs to 78 imported.}

These venues thus state which repository a paper \emph{is about}, rather than which tools it \emph{uses}, forming the precision-oriented ground truth on which the rest of the paper builds.

\subsection{Generated Datasets}
\label{sec:generated}

\Cref{tab:generated} aggregates the per-venue harvest; once normalised and de-duplicated, it forms a unified table of 4{,}397 $\langle$DOI, repo$\rangle$ pairs.

Once compiled, every unique repository URL in the unified table is submitted to Software Heritage's bulk \emph{Save Code Now} API. This mechanism preserves the referenced code and initiates an archival crawl, making each origin's SWHID resolvable so it can be asserted as property \texttt{P6138} on its corresponding Wikidata item. The pipeline itself only reads Wikidata; edits reach the live graph solely through human-reviewed batches (\Cref{sec:methodology}).

\begin{table}
  \caption{Generated $\langle$DOI, repo$\rangle$ pair datasets by venue (successful pairs; in parentheses, the records each venue was harvested from), and their union. No pair is shared across venues, so de-duplication removes nothing at this stage.}
  \label{tab:generated}
  \centering
  \begin{tabular}{llr}
    \toprule
    Source & Access & Pairs \\
    \midrule
    JOSS         & JSON catalogue API                    & 3{,}545~(of 3{,}596) \\
    SoftwareX    & Crossref (ISSN 2352-7110)             & 486~(of 2{,}326) \\
    IPOL         & landing-page BibTeX                   & 275~(of 295) \\
    SIGMOD (ARI) & reproducibility report PDFs (2020--2024) & 91~(of 102 with reports) \\
    \midrule
    Unified      & de-duplicated union                   & \textbf{4{,}397} \\
    \bottomrule
  \end{tabular}
\end{table}

\section{Two Application Profiles: Scholarly Articles and Software Instances}
\label{sec:ontology}

\begin{figure}
  \centering
  \resizebox{\linewidth}{!}{\begin{tikzpicture}
\setlength{\hgap}{4cm}

\node[kgArtClassNode,anchor=center,minimum width=6.6cm] (art) {
  \centering\scshape Scholarly Article
  \nodepart{two}
  \begin{tabular}{@{}l@{~~}l@{}}
    \gelab{P31}\,instance of          & \kt{item}\\
    \gelab{P1476}\,title              & \kt{literal}\\
    \gelab{P356}\,DOI                 & \kt{literal}\\
    \gelab{P1433}\,published in       & \kt{item}\\
    \gelab{P123}\,publisher           & \kt{item}\\
    \gelab{P577}\,publication date    & \kt{literal}\\
    \gelab{P407}\,language of work    & \kt{item}\\
    \gelab{P2093}\,author name string & \kt{literal, ord.\ P1545}\\
  \end{tabular}
};

\node[kgSwClassNode,anchor=center,minimum width=6.6cm,right=\hgap of art] (sw) {
  \centering\scshape Software
  \nodepart{two}
  \begin{tabular}{@{}l@{~~}l@{}}
    \gelabsw{P31}\,instance of        & \kt{item}\\
    \gelabsw{P1324}\,source code repo. & \kt{literal, q.\ P8423/P10627}\\
    \gelabsw{P6138}\,SWH identifier    & \kt{literal}\\
  \end{tabular}
  \nodepart{three}
  \begin{tabular}{@{}l@{}}
    \mo{\itshape descriptive (modelled):}\\
    \mo{P356 $\cdot$ P577 $\cdot$ P348 $\cdot$ P178 $\cdot$ P275 $\cdot$ P277}\\
  \end{tabular}
}
  edge[kgArrIn,bend left=18] node[kgLab] {\gelab{P921 main subject} \kt{(1..*)}} (art)
  edge[kgArrIn,bend left=36,densely dashed] node[kgLab] {\gelabmod{P2860 cites work} \kt{(0..*)}} (art)
  edge[kgArrOut,bend right=22] node[kgLab] {\gelabsw{P1343 described by source} \kt{(1..*)}} (art);

\end{tikzpicture}}
  \caption{Cross-profile intersection between the Scholarly Article Profile (blue, grounded in Wikidata) and the Software Instance Profile (orange, grounded in Wikidata), showing the relational properties linking publications and code repositories. Solid arrows denote relations populated by this import; the dashed grey arrow denotes \emph{cites work} (\texttt{P2860}), which is modelled but left for future work.}
  \label{fig:ontologies_intersection}
\end{figure}
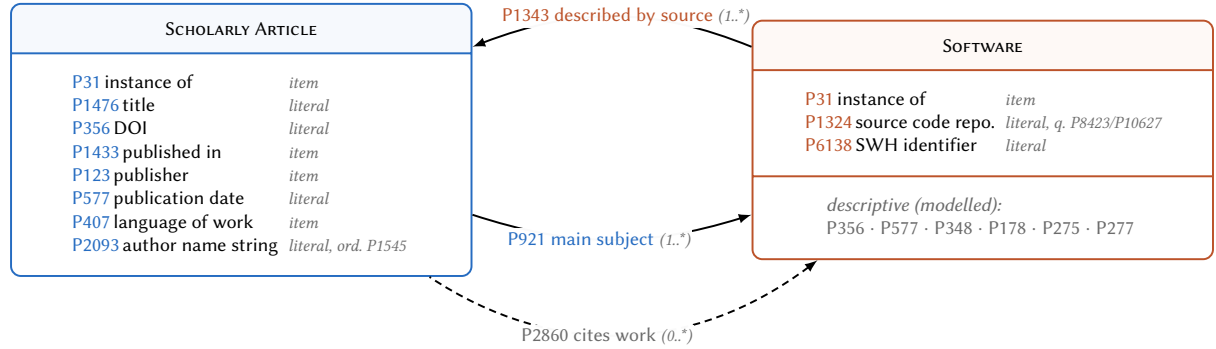

A software-citation link intersects two fundamentally distinct entities: a scholarly article and a software system. Rather than compressing them into a single record, we model each side with a dedicated application profile and connect them via explicit relations (\Cref{fig:ontologies_intersection}). This separation keeps the bibliographic identity of a paper distinct from the archival identity of the code it describes, allowing the reconciliation pipeline to determine the structural depth of each link dynamically.

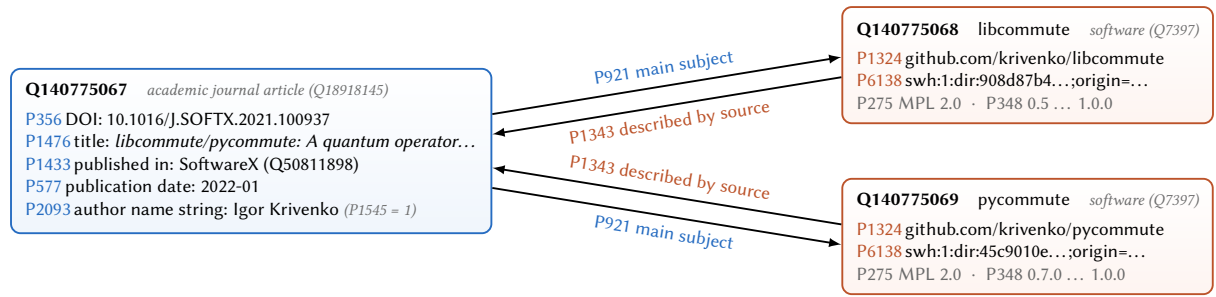
\begin{figure}
  \centering
  \resizebox{\linewidth}{!}{\begin{tikzpicture}

\node[kgArtNode,align=left,inner sep=6pt] (art) {
  {\bfseries Q140775067}\quad\kt{academic journal article (Q18918145)}\\[3pt]
  \gelab{P356}\,DOI: 10.1016/J.SOFTX.2021.100937\\
  \gelab{P1476}\,title: \emph{libcommute/pycommute: A quantum operator\ldots}\\
  \gelab{P1433}\,published in: SoftwareX (Q50811898)\\
  \gelab{P577}\,publication date: 2022-01\\
  \gelab{P2093}\,author name string: Igor Krivenko \kt{(P1545 = 1)}
};

\node[kgSwNode,align=left,inner sep=6pt,anchor=south west] (lib) at ($(art.east)+(5.2cm,0.4cm)$) {
  {\bfseries Q140775068}\quad libcommute\quad\kt{software (Q7397)}\\[3pt]
  \gelabsw{P1324}\,github.com/krivenko/libcommute\\
  \gelabsw{P6138}\,swh:1:dir:908d87b4\ldots;origin=\ldots\\
  \mo{P275 MPL~2.0\enspace$\cdot$\enspace P348 0.5 \ldots\ 1.0.0}
};

\node[kgSwNode,align=left,inner sep=6pt,anchor=north west] (py) at ($(art.east)+(5.2cm,-0.4cm)$) {
  {\bfseries Q140775069}\quad pycommute\quad\kt{software (Q7397)}\\[3pt]
  \gelabsw{P1324}\,github.com/krivenko/pycommute\\
  \gelabsw{P6138}\,swh:1:dir:45c9010e\ldots;origin=\ldots\\
  \mo{P275 MPL~2.0\enspace$\cdot$\enspace P348 0.7.0 \ldots\ 1.0.0}
};

\coordinate (a1) at ($(art.east)+(0,0.55cm)$);
\coordinate (a2) at ($(art.east)+(0,0.25cm)$);
\coordinate (a3) at ($(art.east)+(0,-0.25cm)$);
\coordinate (a4) at ($(art.east)+(0,-0.55cm)$);
\draw[kgArr,-latex] (a1) -- node[sloped,above,font=\sffamily\footnotesize] {\gelab{P921 main subject}} ($(lib.west)+(0,0.12cm)$);
\draw[kgArr,-latex] ($(lib.west)+(0,-0.18cm)$) -- node[sloped,below,font=\sffamily\footnotesize] {\gelabsw{P1343 described by source}} (a2);
\draw[kgArr,-latex] ($(py.west)+(0,0.18cm)$) -- node[sloped,above,font=\sffamily\footnotesize] {\gelabsw{P1343 described by source}} (a3);
\draw[kgArr,-latex] (a4) -- node[sloped,below,font=\sffamily\footnotesize] {\gelab{P921 main subject}} ($(py.west)+(0,-0.12cm)$);

\end{tikzpicture}}
  \caption{Running example: SoftwareX article \texttt{10.1016/j.softx.2021.100937} as imported into Wikidata, one article item and two software items joined by \texttt{P921} and \texttt{P1343} (statements abridged). A third-party bot (\emph{Github-wiki-bot}) added the grey statements afterwards.}
  \label{fig:running_example}
\end{figure}

\paragraph{Running example} SoftwareX article \texttt{10.1016/j.softx.2021.100937} presents two related packages, \texttt{libcommute} and \texttt{pycommute}. Under our profiles it becomes three linked items (\Cref{fig:running_example}): an \emph{academic journal article} (\texttt{Q140775067}) carrying DOI, title, venue, and one \emph{main subject} (\texttt{P921}) statement per package, and two \emph{software} items (\texttt{Q140775068}, \texttt{Q140775069}\footnote{Query: \url{https://w.wiki/UjdF}.}), each with its repository URL (\texttt{P1324}), SWHID (\texttt{P6138}), and a \emph{described by source} (\texttt{P1343}) back-link. Within hours of their creation, a third-party bot added licence (\texttt{P275}) and version (\texttt{P348}) statements to both software items, showing how standalone software nodes accrue metadata (\Cref{sec:granularity}).

Both profiles reuse existing Wikidata classes and properties, a deliberate choice given that engineering new properties is a notorious bottleneck in collaborative knowledge graphs~\cite[\S3.7]{kg_role_dagstuhl}. They are aligned with established vocabularies: schema.org for articles, and the CodeMeta exchange schema~\cite{codemeta} for software.

\subsection{The Scholarly-Article Profile}

A publication is modelled as a Wikidata item instantiating \emph{academic journal article} (\texttt{Q18918145}), a subclass of the \emph{scholarly article} (\texttt{Q13442814}) type applied extensively by the WikiCite initiative~\cite{wikicite_report_2016}; we prefer the narrower class for the journal venues (SoftwareX, JOSS, IPOL), reserving the sibling \emph{conference paper} class (\texttt{Q23927052}) for the SIGMOD conference proceedings. SIGMOD spans both types: papers in the journal \emph{Proceedings of the ACM on Management of Data} take the journal-article class and are \emph{published in} \texttt{Q130602410}, as in the pre-existing WikiCite records, whereas papers in the 2020--2022 conference volumes take the conference-paper class and no \emph{published in} statement, since those volumes have no Wikidata item and we mint no placeholder; for SIGMOD we therefore also record the \emph{publisher} (\texttt{P123}, ACM) on new articles. The article model populates the bibliographic properties of \Cref{tab:artontology}, recording authors as name strings (\texttt{P2093}) with a series ordinal (\texttt{P1545}) rather than as person items. A relational property, \emph{main subject} (\texttt{P921}), bridges each software paper to its software target; a companion property, \emph{cites work} (\texttt{P2860}), models the reference-level \emph{uses} relationship and is reserved for future work over software-citation datasets (\Cref{sec:linking}).

\begin{table}
  \small
  \caption{The scholarly-article profile and its crosswalk to schema.org. Upper block: properties populated by the current import. Lower block: modelled for the reference-level \emph{uses} relationship (future work).}
  \label{tab:artontology}
  \centering
  \begin{tabular}{lll}
    \toprule
    Wikidata property & schema.org & Role \\
    \midrule
    \multicolumn{3}{l}{\small\textit{Populated}} \\
    \texttt{P31}   instance of      & \texttt{@type}                    & article class \\
    \texttt{P1476} title            & \texttt{name} / \texttt{headline} & article title \\
    \texttt{P356}  DOI              & \texttt{identifier} (DOI)         & citable identifier \\
    \texttt{P1433} published in     & \texttt{isPartOf}                 & venue (journal / proc.) \\
    \texttt{P577}  publication date & \texttt{datePublished}            & publication date \\
    \texttt{P123}  publisher        & \texttt{publisher}                & publisher (SIGMOD vols.) \\
    \texttt{P407}  language of work & \texttt{inLanguage}               & language \\
    \texttt{P2093} author name str. & \texttt{author}                   & authorship (ordinal \texttt{P1545}) \\
    \texttt{P921}  main subject     & \texttt{about}                    & subject (the software) \\
    \midrule
    \multicolumn{3}{l}{\small\textit{Modelled (future work)}} \\
    \texttt{P2860} cites work       & \texttt{citation}                 & software the paper \emph{uses} \\
    \bottomrule
  \end{tabular}
\end{table}

\subsection{The Software-Instance Profile}

A software system is represented as an instance of \emph{software} (\texttt{Q7397}), or a specific subclass like \emph{free software} (\texttt{Q341}) when supported by evidence. Its properties serve two distinct roles. \emph{Identifying} properties establish cross-archive identity and serve as the primary join keys for reconciliation; the current import populates the source code repository (\texttt{P1324}, qualified by its version control system \texttt{P8423} and web interface software \texttt{P10627}) and the SWHID (\texttt{P6138}), together with the back-link \emph{described by source} (\texttt{P1343}) to the parent paper. \emph{Descriptive} properties, which the model provides to enrich an item once its identity is resolved, are populated as such evidence accrues: an archival deposit DOI (\texttt{P356}) when an artefact is registered on a platform like Zenodo, release date (\texttt{P577}), software version (\texttt{P348}), developer (\texttt{P178}), copyright license (\texttt{P275}), and programming language (\texttt{P277}). \Cref{tab:swontology} illustrates this schema and its crosswalk to CodeMeta 3.0, whose terms come from schema.org except \texttt{referencePublication}, which is CodeMeta's own.

This descriptive schema refines the repository-centric model Turki et al.~\cite{rasberry_github_wikidata_2023} used to track GitHub repositories for Wikimedia tooling. Whereas they harvested transient platform metrics (such as star counts, forks, and last-push dates) across a single forge, our model foregrounds three descriptive properties core to citable identity, to be populated as evidence accrues: programming language (\texttt{P277}), license (\texttt{P275}), and developer (\texttt{P178}). We deliberately store both the repository URL (\texttt{P1324}) and the SWHID (\texttt{P6138}), since the two play complementary roles. The URL points to where development currently happens and may change when a project migrates or is renamed. In contrast, the SWHID is a permanent, content-addressed snapshot minted uniformly across forges, decoupling identity from any hosting platform and resisting link rot.

\begin{table}
  \small
  \caption{The software-instance profile and its crosswalk to CodeMeta 3.0. Upper block: identity and cross-link properties populated by the current import (\texttt{P1324} qualified by version control system \texttt{P8423} and web interface software \texttt{P10627}). Lower block: descriptive properties the model provides, populated as evidence accrues. \emph{Voc.} records the namespace each term resolves to in the CodeMeta context: \texttt{sdo} for \texttt{schema:} (schema.org), \texttt{cm} for CodeMeta's own \texttt{codemeta:}; \texttt{@type} is a JSON-LD keyword.}
  \label{tab:swontology}
  \centering
  \begin{tabular}{llcl}
    \toprule
    Wikidata property & CodeMeta term & Voc. & Role \\
    \midrule
    \multicolumn{4}{l}{\small\textit{Populated (identity + link)}} \\
    \texttt{P31}   instance of          & \texttt{@type}                & ---           & software class \\
    \texttt{P1324} source code repo.    & \texttt{codeRepository}       & \texttt{sdo}  & repository URL (identity) \\
    \texttt{P6138} SWH identifier       & \texttt{identifier} (SWHID)   & \texttt{sdo}  & archival anchor (identity) \\
    \texttt{P1343} described by source  & \texttt{referencePublication} & \texttt{cm}   & back-link to describing paper \\
    \midrule
    \multicolumn{4}{l}{\small\textit{Modelled (descriptive enrichment)}} \\
    \texttt{P356}  DOI                  & \texttt{identifier} (DOI)     & \texttt{sdo}  & software's own deposit DOI \\
    \texttt{P577}  publication date     & \texttt{datePublished}        & \texttt{sdo}  & release / version date \\
    \texttt{P348}  software version id. & \texttt{softwareVersion}      & \texttt{sdo}  & version label \\
    \texttt{P178}  developer            & \texttt{author}               & \texttt{sdo}  & authorship \\
    \texttt{P275}  copyright license    & \texttt{license}              & \texttt{sdo}  & license \\
    \texttt{P277}  programmed in        & \texttt{programmingLanguage}  & \texttt{sdo}  & implementation language \\
    \bottomrule
  \end{tabular}
\end{table}

\subsection{Linking the Two Profiles}
\label{sec:linking}

The profiles intersect via directed relationships rather than shared identifier properties. From the article side, \emph{main subject} (\texttt{P921}) indicates that a dedicated software paper is structurally \emph{about} a software item, whereas \emph{cites work} (\texttt{P2860}) models an ordinary citation where a paper simply \emph{uses} a tool, a case we populate only from the software-mention datasets, left to future work. Conversely, the software item uses \emph{described by source} (\texttt{P1343}) as an inverse back-link to its primary paper. This cross-profile bridge (\Cref{fig:ontologies_intersection}) is not strictly one-to-one, since a paper presenting several packages carries one \texttt{P921} per software item (\Cref{fig:running_example}).

This also resolves a common confusion about DOIs: on an article item, \texttt{P356} identifies the publication, whereas on a software item it identifies an independent archival deposit (such as a Zenodo release). The two never coincide, avoiding the conflation common in single-record metadata schemas.

\subsection{Two Levels of Reconciliation Granularity}
\label{sec:granularity}

Reconciling an SWH object with Wikidata can proceed at two granularities. At the coarser \emph{reference level}, no dedicated software item is minted: the article item is retrieved or created via its DOI, and the SWHID or repository URL is attached directly as a statement reference. This keeps archived source code discoverable from the bibliographic graph without full entity resolution, pointing to any layer of the Merkle graph down to individual files or revisions (\Cref{fig:granularity}). It degrades gracefully when metadata is sparse, making it ideal for the long tail of incidental software mentions.

At the finer \emph{node level}, software is promoted to a first-class Wikidata item (\texttt{Q7397}) carrying its own identity properties (\texttt{P1324}, \texttt{P6138}) and descriptive statements (\texttt{P348}, \texttt{P275}, \texttt{P277}). The item is linked to the publication via \texttt{P921} on the article (reserving \texttt{P2860} for reference-level usage) and \texttt{P1343} on the software. This tier enables rich semantic queries (e.g., listing all papers citing a tool or mapping software licences across a discipline) but requires strict disambiguation to prevent duplicate nodes. Following the 2025 query-service split that isolates articles and software across separate SPARQL endpoints~\cite{wdqs-split}, our cross-entity properties (\texttt{P921}, \texttt{P1343}) provide explicit keys for efficient federated joins on WDQS, while QLever offers an alternative endpoint.\footnote{\url{https://qlever.dev/wikidata/}}

Node level is our default for software that is the explicit subject of a paper, as in our venues. Articles and code routinely carry different licences (e.g., CC~BY for a paper, MIT for its code), so \texttt{P275} needs a dedicated software node, and standalone nodes accrue versions and statements without later disentangling reference links (\Cref{fig:running_example}). The reference level is reserved for incidental mentions.

\subsection{Mapping Software Heritage Objects}

Node-level identity mirrors the Merkle graph topology. Each SWH \emph{origin}, defined by its upstream repository URL, maps to a single software item, with the URL serving as the \texttt{P1324} value and the primary reconciliation key. Snapshots and revisions beneath this origin capture version-level identity: the SWHID of a tagged release or snapshot is populated as a \texttt{P6138} value, which the model further qualifies with a version string (\texttt{P348}) and release date (\texttt{P577}) where available. Because citations generally target the software concept rather than an isolated commit, we anchor node identity at the origin level, recording fine-grained SWHIDs as qualified, version-scoped statements to control concept-versus-version ambiguity.

Bibliographic and identifier statements carry a structured reference block capturing the source corpus, evidence URL or DOI, and extraction date.

\section{The Reconciliation Pipeline}
\label{sec:methodology}

The pipeline is the core component of an open-source library released alongside our harvested datasets for full reproducibility.\footnote{Code, data, and import batch reports: \url{https://github.com/ftosoni/swh-wd-reconciliation}.} It executes five sequential stages (harvesting, bulk archiving, lookup, statement generation, and a reviewed two-pass import), following knowledge-engineering best practices by starting from explicit use cases and enforcing strict provenance for every claim~\cite[\S5.8]{kg_role_dagstuhl}.

\textit{Harvesting} normalises heterogeneous metadata into a uniform table of $\langle$DOI, repo-URL$\rangle$ pairs. A tiered API strategy across three software journals and SIGMOD ARI reports (\Cref{sec:groundtruth}) yields 4{,}397 pairs (\Cref{tab:generated}).

\textit{Bulk archiving} normalises harvested URLs, filters out non-code hosts, and discards existing SWH links, retaining only origins on a code-hosting allowlist (\texttt{CODE\_HOST}). This filters out generic homepages, docs, and direct downloads. The remaining origins are submitted to SWH's bulk \emph{Save Code Now} API, yielding 4{,}086 unique repositories to archive from the initial 4{,}397 pairs. \textit{SWHID retrieval} then queries visit histories, extracts the crawl closest to publication, and resolves 4{,}244 pairs to a qualified \texttt{P6138} SWHID value (with the remainder retaining repository URLs).

\textit{Lookup} decides, for each pair, whether to create or enrich. In our experience, the only reliable identity keys are exact identifiers: an article is matched by its DOI (\texttt{P356}, case-insensitively) and a software item by its normalised repository URL (\texttt{P1324}), with the SWHID (\texttt{P6138}) as a secondary key (\Cref{sec:evaluation}). Name-based matching is never used to merge items, since software labels are short and heavily reused (e.g., two unrelated projects named \texttt{umami}). We create missing nodes, while enriching existing ones rather than duplicating them. Pairs that would collapse distinct entities onto one node, such as a repository shared by several tools or a publisher identifier collision (\Cref{sec:groundtruth}), are excluded after manual review rather than merged, preventing item conflation and duplication~\cite{pellizzari2023conflations}.

\textit{Statement generation} turns the validated pairs into CSV tables, imported through OpenRefine or converted into QuickStatements commands, with target QIDs (or new-item placeholders), properties, values, and provenance blocks. Article titles are stripped of JATS/MathML markup and HTML entities, with whitespace collapsed to create clean labels and \texttt{P1476} values. Repository URLs serving as \texttt{P1324} values are trimmed of sub-paths (e.g., \texttt{/tree/}, \texttt{/src/}, or \texttt{/issues}) back to the repository root. To keep label and description jointly unique, software items with ambiguous names (such as the two \texttt{umami} projects) append the repository slug to their description. Granularity choices follow the policy in \Cref{sec:granularity}.

\textit{Import} runs in two passes, because a batch cannot use an item it creates as the value of another statement. The first pass creates the missing software and article items (via OpenRefine or QuickStatements~3.0); the second, fed with the QIDs listed in the first pass's run report, completes the \texttt{P921}/\texttt{P1343} cross-links and adds the author strings. The authors inspected and ran every batch, creating 4{,}182 software and 2{,}326 article items. Our code repository lists every batch (also browsable on EditGroups) and includes the QuickStatements run reports; once imported, the statements become part of Wikidata's public RDF graph, queryable live through its SPARQL endpoints and redistributable through its RDF dumps.

\section{Evaluation}
\label{sec:evaluation}

We evaluate the pipeline's \emph{lookup} stage, which, for each harvested node, determines whether an equivalent item already exists in Wikidata by querying the live graph directly. This both quantifies the enrichment opportunity and prevents duplicate nodes from being created before any edit is staged.

A read-only query against the live Wikidata graph (retrieved 29 July 2026) resolves each of the 4{,}397 $\langle$article, software$\rangle$ pairs from the four venues against existing items.\footnote{The exact read-only SPARQL \texttt{SELECT}s (indexing \texttt{P1324}/\texttt{P6138} for software on the main endpoint and matching \texttt{P356} for articles on the scholarly subgraph) are part of the released code (\texttt{precheck\_wikidata.py}); none writes to Wikidata.} A software node is matched by its source code repository URL (\texttt{P1324}), with the SWHID (\texttt{P6138}) as a secondary key; an article node is matched by its DOI (\texttt{P356}). Repository URLs are canonicalised on both sides before comparison (lower-casing the whole URL and stripping the scheme, a leading \texttt{www.}, trailing slashes, and \texttt{.git} suffixes). \Cref{tab:precheck} breaks the outcome down by venue: only 82 of the 4{,}397 software repositories (1.9\%) already have a Wikidata item, so 4{,}315 are candidates for new software nodes, whereas 1{,}967 (44.7\%) of the articles already exist through WikiCite. These counts span the entire harvested corpus, before the per-venue editorial refinements in \Cref{sec:groundtruth}, so they measure the enrichment opportunity rather than any single import batch.

\begin{table}
  \caption{Lookup outcome against the live Wikidata graph (read-only, 29 July 2026). Software nodes are matched by repository URL (\texttt{P1324}); articles by DOI (\texttt{P356}).}
  \label{tab:precheck}
  \centering
  \begin{tabular}{lrrrrr}
    \toprule
    & & \multicolumn{2}{c}{Software} & \multicolumn{2}{c}{Article} \\
    \cmidrule(lr){3-4}\cmidrule(lr){5-6}
    Venue & Pairs & exist. & new & exist. & new \\
    \midrule
    SoftwareX & 486 & 6 & 480 & 210 & 276 \\
    JOSS      & 3{,}545 & 76 & 3{,}469 & 1{,}691 & 1{,}854 \\
    IPOL      & 275 & 0 & 275 & 31 & 244 \\
    SIGMOD    & 91 & 0 & 91 & 35 & 56 \\
    \midrule
    \textbf{Total} & \textbf{4{,}397} & \textbf{82} & \textbf{4{,}315} & \textbf{1{,}967} & \textbf{2{,}430} \\
    \bottomrule
  \end{tabular}
\end{table}

Because venue policies fix the coupling between a DOI and its repository, the corpus also gives an objective precision baseline: a mapping that pairs a venue DOI with a different repository is incorrect. Computing structural precision metrics on the full corpus is ongoing work.

A primary limitation of the current prototype is that repository matching relies on URL equality after canonicalisation rather than graph traversal within SWH. Canonicalisation absorbs superficial variants, but the match still produces false negatives when projects migrate between hosting platforms (e.g., from Bitbucket to GitHub) without corresponding updates to their Wikidata \texttt{P1324} values. Anchoring identity directly on the SWHID, which remains invariant across code migrations, is our primary planned technical improvement.

\paragraph{Lessons learned} Most defects arose during extraction rather than modelling, and manual audits caught them all before upload. They took three forms: publisher metadata pointing to a sibling record (the SoftwareX identifier collision); links corrupted in transit, such as URLs broken by PDF line-wrapping in SIGMOD reports or concatenated SWHIDs in IPOL exports; and repositories that break the one-repository-per-item assumption, being shared by several tools or holding only the paper (\Cref{sec:groundtruth}). We therefore recommend that venues publish the software link (repository URL or SWHID) as structured metadata, as JOSS and IPOL already do through their catalogue and BibTeX exports and ideally in the DOI record itself, rather than as free text in a PDF or a publisher-side mirror.

\section{Conclusion}
\label{sec:conclusion}

We presented an end-to-end pipeline that reconciles software between Software Heritage and Wikidata over 4{,}397 editorially verified $\langle$DOI, repository$\rangle$ pairs, of which only 82 repositories were already in Wikidata. Its contributions are threefold: dual application profiles that map SWH concepts onto Wikidata while keeping publications and software distinct at the reference or node level; an input format that natively accepts COAR Notify payloads; and reviewable, provenance-tagged upload batches that prevent unvetted writes to the live graph.

Immediate next steps involve measuring stage-by-stage precision, connecting the backend to a live COAR Notify stream, and extending harvesting to venues that mandate artefact links, such as the Resource tracks of ISWC and ESWC. Finally, adding multilingual labels and descriptions remains essential to maximise Wikidata's utility as a cross-lingual hub.

\begin{acknowledgments}

  This study was funded by Wikimedia Italia through its micro-grant programme (\url{https://w.wiki/UHgf}). Francesco Tosoni was funded by the Alfred P. Sloan Foundation with the grant \#\href{https://sloan.org/grant-detail/g-2025-25193}{G-2025-25193} (\url{sloan.org}).

\noindent The authors thank the anonymous reviewers for their constructive comments and Morane Gruenpeter for helpful discussions.

\noindent \Cref{fig:coar-notify-model} adapts the SoFAIR workflow of Cancellieri et al.~\cite{coar} (CC~BY~4.0) and uses Font Awesome Free icons (CC~BY~4.0). All logos are trademarks of their respective owners.

\end{acknowledgments}

\section*{Declaration on Generative AI}
During the preparation of this work, the authors used Claude (Anthropic) to draft content, paraphrase, reword, and generate images, and Grammarly to improve writing style. In more detail, the authors used Claude (Anthropic) to rewrite their data import logs and reconciliation notes in discursive form to produce a preliminary draft of the corresponding sections, shorten and reword text, and write the TikZ code for the figures. The released code was also written with assistance from Claude (Anthropic) and reviewed through a manual code-review process. After using these tools, the authors reviewed and edited the content as needed, checked every citation against the cited works, and take full responsibility for the publication's content.

\bibliography{references}

\end{document}